\documentclass[twocolumn,showpacs,prl]{revtex4}

\usepackage{graphicx}
\usepackage{dcolumn}
\usepackage{bm}

\begin{document}

\setlength{\baselineskip}{0.4cm}
\addtolength{\topmargin}{1.5cm}

\title{Quasiperiodic events in an earthquake model}

\author{O. Ramos$^{1}$, E. Altshuler$^{2}$, and K. J. M{\aa}l{\o}y$^{1}$}

\affiliation{ $^1$Department of Physics, University
of Oslo, P.O.B. 1048, Blindern N-0316, Oslo, Norway\\
$^2$``Henri Poincar\'e" Group of Complex Systems, Physics Faculty,
University of Havana, 10400 Havana, Cuba }

\date{\today}
\begin{abstract}

We introduce a modification of the OFC earthquake model [Phys. Rev. Lett. 68,
1244 (1992)] in order to improve resemblance with the Burridge and
Knopoff mechanical model and with possible laboratory experiments.
A constant force continually drives the system, and thresholds are
distributed randomly following a narrow distribution. We find
quasiperiodic behavior  in the avalanche time series with a period
proportional to the degree of dissipation of the system.
Periodicity is not as robust as criticality when the threshold
force distribution widens; and foreshocks and aftershocks are
connected to the observed periodicity.

\end{abstract}


\maketitle

The concept of Self-organized criticality (SOC) \cite{Bak et
al-1987_88}, introduced by Bak, Tang and Wiesenfeld in 1987, was
an attempt to explain the appearance of scale invariance in
nature. In their sandpile model both the random, slow addition of
``blocks" on a two dimensional lattice and a simple, local, and
conservative rule drive the system into a critical state where
power law distributed avalanches maintain a steady regime far from
equilibrium. There is no correlation between the avalanches, and
eventually they reach the boundaries of the lattice liberating the
excess of energy. Five years later Olami, Feder and Christensen
(OFC) made an important contribution to the SOC ideas by mapping
the Burridge and Knopoff spring-block model \cite{Burridge and
knopoff-1967} into a nonconservative cellular automata
\cite{OFC-1992, Christensen and Olami -1992}, simulating the
earthquake's behavior and introducing dissipation in the family of
SOC systems. The fact that avalanches are uncorrelated in the
sandpile model has been used as an argument to propose that it is
not possible to predict real earthquakes \cite{Geller et al
-1997}. However, foreshocks, aftershocks and clustering properties
\cite{Gerstenberger et al -2005} indicate the existence of
correlation between different events. Many seismologists believe
that large earthquakes are quasiperiodic \cite{Bakun and Lindh
-1985, Savage and Cockerham -1987}, but periodic behavior has
appeared in theoretical models only as a special or as a trivial
solution \cite{Carlson and Langer -1989, Rice -1993, Xu and
Knopoff -1994}.

 The spring-block model consists in a two dimensional array of
blocks on a flat surface. Each block is connected with its four
nearest neighbors, and in the vertical direction, to a driving plate
which moves horizontally at velocity $v$. The connections are made
by springs, and when the force acting on a block overcomes the
static friction with the surface, the block slips. Then a
redistribution of forces takes place in the neighbors that
eventually trigger new displacements. In the OFC model the force on
a block is stored in a site of a squared lattice, and the static
friction threshold has the same value for all blocks. Starting from
a random distribution of forces, the site closest to the threshold
is found and the {\it exact} force necessary to provoke a slip in
this block is added to every site of the grid. This infinitely
accurate tuning is only possible in the mechanical model if the
displacement of the plate is infinitely slow ($v\rightarrow0$),
considering the fact that the real time resolution is finite. When a
site reaches the threshold it is set to zero, and a fraction
$\alpha$ of its force is redistributed to its neighbors. If
$\alpha=1/4$ the system is conservative. If one of the neighbors
reaches the threshold the process is repeated until all the sites
have their values below the threshold. The number of slips (or the
number of sites involved in the process) is defined as the avalanche
size. Then again, the site closest to the threshold is found and a
new avalanche is triggered. The avalanche distributions follow power
laws and by varying the degree of dissipation $\alpha$, the slope of
the distribution can be tuned. Although the criticality of the
system in the nonconservative regime ($\alpha<1/4$) has been widely
debated \cite{Carvalho and Prado -1992, Miller and Boulter -2002},
the model results in a power law distribution of avalanches (for
$\alpha\sim0.2$) similar to the Gutenberg-Richter law
\cite{Gutenberg and Richter -1956} and also reproduces other
characteristics of real earthquakes \cite{Hergarten and Neugebauer
-2002, Peixoto and Prado -2004}.

In the aim of improving resemblance with the mechanical model we
have introduced two variations in the OFC model. First: Thresholds
are distributed randomly following a Gaussian distribution of
standard deviation $\sigma$. When a block slips a new threshold is
imposed to its site. This is closer to the actual block-surface
friction problem and allows the system to start from a configuration
of zero force in every block \cite{comment}. Second: Instead of
assuming infinitely accurate tuning, we add a quantum of force in
each step. In the mechanical model this is equivalent to a finite
velocity (considering finite time resolution). The dynamics become
more complex because several sites can reach the threshold as the
plate moves, so several clusters start to grow and eventually they
can touch each other merging into a single one. The avalanche size
is defined as the number of slips (or number of sites involved) in
each cluster.

\begin{figure}[t]
\vspace{-1cm}
\includegraphics[height=2.912in, width=3.7in]{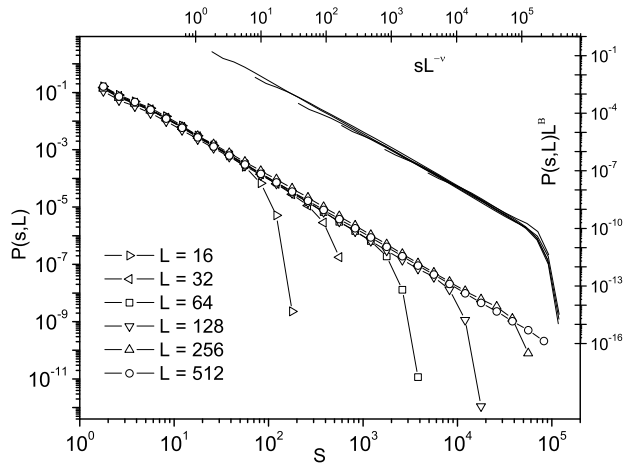}
\vspace{-0.3cm} \caption{\label{fig:f1} Avalanche size
distributions for the number of slips in each cluster, with
$\alpha = 0.2$ and $\sigma = 0.001$. The slope is $-1.91$. Inset:
collapse of all the curves under the scaling relation
$P(s,L)L^{\beta}=f(sL^{-\nu})$, with $\beta = 4.2$ and $\nu =
2.2$.}
\end{figure}

\begin{figure}[b]
\vspace{-0.3cm}
\includegraphics[height=2.746in, width=3.3in]{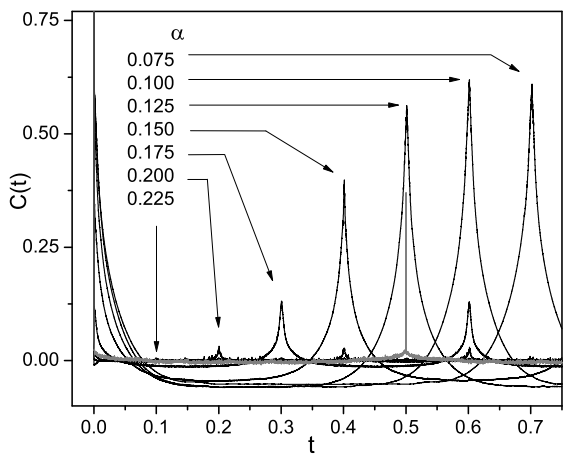}
\vspace{-0.3cm} \caption{\label{fig:f2} Auto-correlation function
for the avalanche time series (eq:1) for a system with
$\sigma=0.001$ and L=128. Notice in gray the same function for a
system with L=128, $\alpha=0.125$, but with $\sigma=0$.}
\end{figure}

In order to study the temporal series of avalanches we need to
define a unit of time. Let us consider the case $\alpha=0$,
$\sigma=0$ (isolated blocks with single threshold). When the force
that drives the system reach a value $F_m$ equal to the threshold,
every site returns to its original value and a trivial periodicity
rules the dynamics; this period is a natural unit. Therefore our
unit of time is the real time $T_m$ that a mechanical model spends
to add a force $F_m$ to the system.

We performed simulations taking a quantum of force $\delta
F=10^{-4}$ and threshold values around 1 for different $\sigma$
values: 0, 0.001, and 0.01. Then an isolated block would need $10^4$
steps to reach the threshold, so it would spend a time $\delta
t/T_m=10^{-4}$ in each step. We find that the avalanche size
distributions follow power laws for different values of $\alpha$,
with cut-offs sensitive to the size of the system. The avalanche
time series show quasiperiodic behavior with a period proportional
to the degree of dissipation (similar behavior can be found in the
OFC model); periodicity is less robust than criticality in the
system, and foreshocks and aftershocks are strongly related with the
observed periodicity. All the simulations presented in this paper
took place with open boundary conditions, but the same results were
obtained in a system with free boundary conditions \cite{Christensen
and Olami -1992}.

Figure 1 shows the avalanche size distribution for different sizes
of the lattice (more than $3\cdot10^8$ clusters for L=512), where
avalanche size is defined as the number of slips in each cluster.
The system, with $\alpha= 0.2$ and $\sigma = 0.001$, has been
driven adding a force $\delta F=10^{-4}$ in each step. The
distributions follow a power law with a slope equal to -1.91. The
curves present cut-offs sensitive to the size of the system but
they collapse (see inset) when the finite size scaling relation
$P(s,L)L^{\beta}=f(sL^{-\nu})$ is applied with $\beta = 4.2$ and
$\nu = 2.2$. This $\nu$ value larger than the system dimension
indicates that for larger avalanches, sites slip more than once.
Simulations for other $\alpha$ values with L=512 displayed that
the absolute value of the slope of the distributions increases as
the dissipation decreases, in contrast to \cite{Lise and Paczuski
-2001}. Also the power law behavior of the avalanche distributions
for small $\alpha$ values is more robust in this finite velocity
model than in the original OFC one due to the fact that larger
clusters have less probability in the OFC model. The avalanche
size distributions (at least for $\alpha>0.1$) do not suffer
considerable variations when $\sigma$ moves from 0 to 0.01. For
$\sigma=0.1$ the system is not critical anymore, but this is not a
realistic value for the fluctuations of the friction force
associated to an interface between a block and a flat surface.
\begin{figure}[b]
\vspace{-0.3cm}
\includegraphics[height=3.88in, width=3.2in]{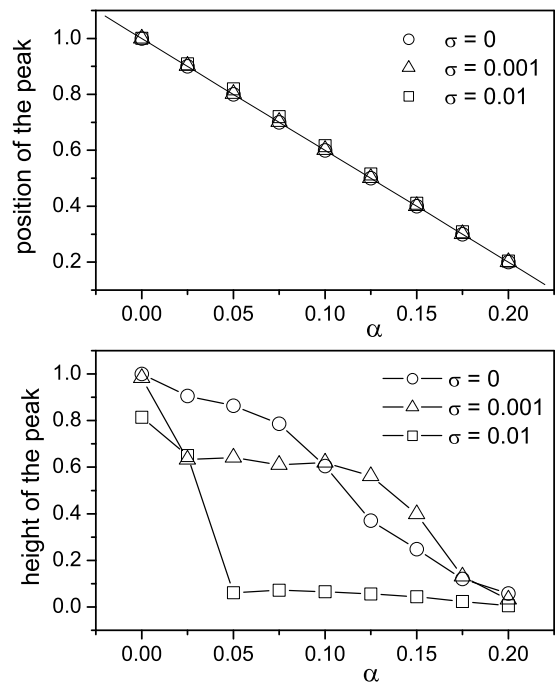}
\vspace{-0.3cm} \caption{\label{fig:f3} above: position of the
peaks for the correlation function of the avalanche time series
(eq:1) for different $\sigma$. They follow the equation
$T=Tm(1-4\alpha)$. below: height of the peaks.}
\end{figure}

The analysis of the auto-correlation function
\begin{equation}
  \label{eq1}
  C(t)=\frac{\sum ( s(\tau)\cdot s(\tau+t))-<s(\tau)>^{2}}{\sum ( s(\tau)-<s(\tau)>)^{2}}
\end{equation}
\noindent where $s(t)$ corresponds to the avalanche time series
(Figures 2 and 3) displays a strong correlation between
avalanches. Some of the peaks for $\sigma=0.001$ are shown in
Figure 2. The position and height of all the peaks for different
$\sigma$ appear in Figure 3. The position of the peaks indicates
that for every analyzed $\sigma$ the system has a quasiperiodic
behavior with a period proportional to the degree of dissipation:
\begin{equation}
  \label{eq2}
  T=T_m(1-4\alpha)
\end{equation}
\noindent In general the height of the peaks decreases as the
dissipation decreases; and for $\alpha=0.225$ the only system where
a sign of periodicity can be seen is for $\sigma=0.001$.
Nevertheless, the peak is just a little higher than the background
noise (see Figure 2). For the conservative case, as in the
``sandpile" model, avalanches are uncorrelated. Periodicity in the
system is not as robust as criticality when $\sigma$ varies: peaks
are almost delta functions for $\sigma=0$ (see curve in gray in
Figure 2) but their width increases dramatically when $\sigma$
increases. The height of the peaks shows a monotonous variation with
$\alpha$ for $\sigma=0$, but for larger $\sigma$ there are local
variations not well understood yet. Although the curves from smaller
$\sigma$ values cross each other, the peaks for larger $\sigma$ are
extremely wide, noisy, and have a very small height, indicating that
periodicity vanishes when noise is added to the system.

\begin{figure}[t]
\vspace{-0.3cm}
\includegraphics[height=2.567in, width=3.3in]{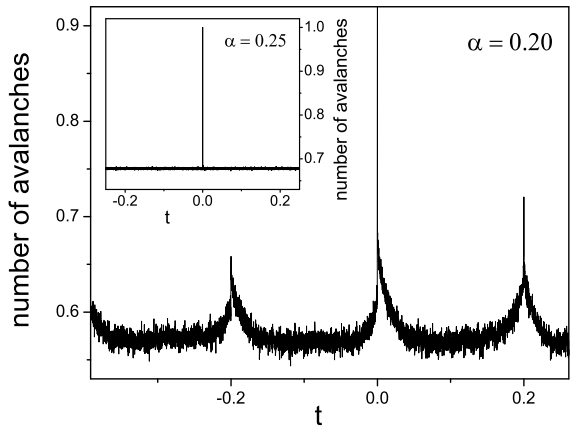}
\vspace{-0.3cm} \caption{\label{fig:f4} Number of avalanches in
the vicinity of a large avalanche, normalized to the total number
of large avalanches, in a system with L=128, $\sigma = 0$ and
$\alpha = 0.2$. Inset: in a system with L=128, $\sigma = 0$ and
$\alpha = 0.25$ (conservative case).}
\end{figure}

\begin{figure}[t]
\vspace{-0.3cm}
\includegraphics[height=2.1in, width=3.5in]{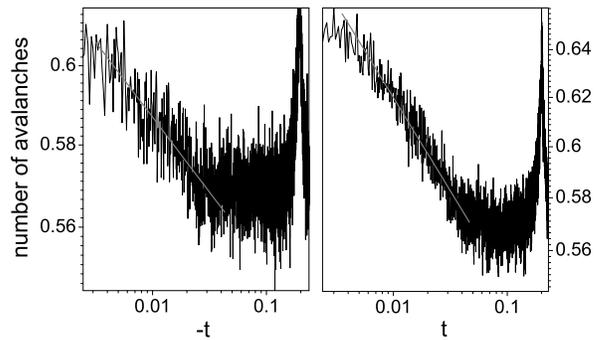}
\vspace{-0.3cm} \caption{\label{fig:f5} Foreshocks (left) and
aftershocks (right) for the avalanches displayed in Figure 4. They
have been plotted in log-log scale for comparative purposes. The
slopes are -0.03 and -0.05 for foreshocks and aftershocks
respectively.}
\end{figure}

\begin{figure}[b]
\vspace{-0.3cm}
\includegraphics[height=2.11in, width=3.5in]{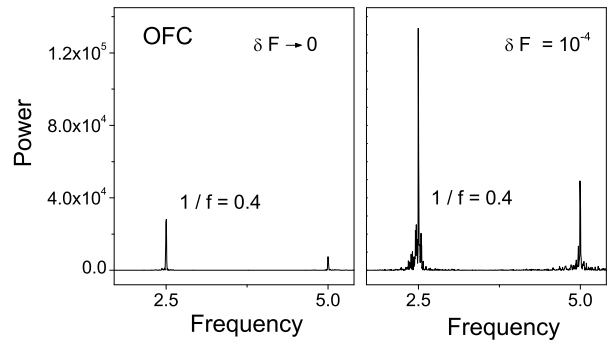}
\vspace{-0.3cm} \caption{\label{fig:f6} Power spectrum for the
avalanche time series for the OFC model (left) and for our model
(right). Both with L=128,  $\alpha=0.15$ and $\sigma=0$.}
\end{figure}

The behavior of the earthquake time series in the vicinity of a
large event have been studied for more than a century, and Omori's
law is one of the most well known issues in seismology: {\it the
rate of aftershocks decays following a power law with slope -1
after a main shock}. We have characterized the avalanches in our
model into small, medium or large in the following way: In the
avalanche size distribution, in a log-log plot, we take the linear
zone; in the case of L=128 it spans from $1$ to $6\cdot10^{3}$
(see Figure 1). Then this interval is divided in three zones
logarithmically equispaced. As a result, avalanches smaller than
$18$ are considered small, those lying between $19$ and $330$ are
medium, and those greater than $330$ are large. The number of
avalanches larger than zero around a large one (normalized to the
total number of large avalanches) for a system with L=128 is
displayed in Figure 4. For $\alpha=0.2$ we see clear signs of
periodicity with a period equal to 0.2, and the decays in the
avalanche rate around a large event are strongly connected with
this observed periodicity. For the conservative case
($\alpha=0.25$) the large avalanches are not preceded nor
succeeded by any event in the avalanche time series. If we plot
both the before and after decays for $\alpha=0.2$ in a log-log
graph (just for comparative purposes) and we fit lines (Figure 5)
the slopes are -0.03 and -0.05 respectively for foreshocks and
aftershocks. If instead of the analysis of the total number of
avalanches around a large shock we do not consider the small ones,
and the same analysis is made, then both slopes are around -0.18.
These small values are in agreement with the previous studies in
the OFC model \cite{Hergarten and Neugebauer -2002}, but due to
the criterion chosen to define a large event, they are sensitive
to the system size.

In order to construct an avalanche time series for the OFC model we
need to define a scale $\delta t$ to measure the time. In the finite
velocity model it is natural to choose $\delta t/T_m = \delta F$, we
will use the same value for the OFC one (the qualitative result is
independent of $\delta t$). If more than one event took place in the
same $\delta t$, the value of the avalanche size in this interval is
equal to the sum of the avalanche size in all those events. The
power spectrum for the avalanche time series for the OFC model with
L=128 and $\alpha=0.15$ appears in Figure 6. It displays a sharp
peak at a frequency equal to 2.5 corresponding to a period equal to
0.4. The power spectrum for a finite velocity model with the same L,
$\alpha$, and $\sigma=0$ shows no qualitative differences with the
OFC one, but the intensity of the peaks are larger. This shows that
periodicity is not confined to this particular finite velocity model
but it is a general characteristic of the OFC one. Nevertheless, in
this case, as $v\rightarrow0$, the unit of time
$T_m\rightarrow\infty$. The same periodicity was found in the
analysis of foreshocks and aftershocks. The correlation function
method did not bring satisfactory results due to poor statistics for
small values of $\alpha$ (the peaks appear for the respective
periods, but the background noise shows large fluctuations for small
$\alpha$ values).

Due to the relation between dissipation and periodicity in the model
we performed a few simulations where the value of $\alpha$ is not
constant but randomly distributed following a Gaussian centered in
$\alpha$, with standard deviations $\sigma_{\alpha}$ equal to
$0.005$, $0.01$, and $0.02$. For $\alpha=0.2$ and the
$\sigma_{\alpha}$ values being 0.005 and 0.01, the height of the
peak in the correlation function decreases more than $90{\%}$ and
the period widens up to $5{\%}$. The avalanche size distributions do
not suffer considerable variations. Similar results were obtained
for $\alpha=0.15$. For $\sigma_{\alpha}$ equal to 0.02 criticality
disappears for both values of $\alpha$. This corroborates that
periodicity is more fragile than criticality when noise is added to
the system.

Quasiperiodic signals in earthquake time series has been used in an
attempt to predict the next main shock, but generally with
unsuccessful results \cite{Bakun and Lindh -1985, Savage and
Cockerham -1987}. This situation, in combination with poor
statistics and the lack of a theory that explains periodicity have
created doubts about the real existence of those series of
quasiperiodic events \cite{Kagan -1996}. Gao {\it et al} \cite{Gao
et al -2000} found an annual periodicity following the 1992 Landers
earthquake in California, but they suggest seasonal differences in
water extraction rates, rainfall and barometric pressure as the
cause of it. Considering the periodicity found in this simple
mapping of the block-spring model into a nonconservative cellular
automata, we can speculate that the earthquake's natural behavior is
a quasiperiodic state and that the variations or absence of
periodicity is due to changes in the dissipative regime and/or in
the relative velocity of the plates and/or in the amount of energy
that can be stored in a given zone between two tectonic plates
(related to our threshold that rules the unit of time).

In conclusion, we have introduced two variations in the OFC model
improving resemblance with the spring-block one, and bridging the
gap between the model and possible experiments. We found
quasiperiodic behavior in the system with a period proportional to
the degree of dissipation. Foreshocks and aftershocks are strongly
connected with the observed periodicity, and for small variations
in the thresholds or in the degree of dissipation, periodicity
tends to vanish, while the system remains showing avalanche size
distributions that follow power laws.

We thank R. Toussaint, J. Schmittbuhl, K. Bassler and S. Santucci
for helpful discussions, and J.S. Chang for a careful reading of
the manuscript. This work was supported by NFR, the Norwegian
Research Council through a Petromax and a SUP grant.


\begin{thebibliography}{99}
\bibitem{comment} In the OFC model randomness is introduced ab
initio, but in the spring-block system it is a consequence of the
fluctuations of the friction force.
\end{thebibliography}
\end{document}